\documentclass[%
 preprint,
 amsmath,amssymb,
 aps, physrev,
]{revtex4-2}

\usepackage{graphicx}% Include figure files
\usepackage{dcolumn}% Align table columns on decimal point
\usepackage{bm}% bold math
\usepackage{csquotes}

\begin{document}

\preprint{}

\title{\textbf{A Thermodynamic Model for Thermomigration in Metals} 
}% 

\author{Daniel J. Long}
 \email{Contact author: daniel.long@eng.ox.ac.uk}
\author{Edmund Tarleton}
\author{Alan C.F. Cocks}
\author{Felix Hofmann}
 \email{Contact author: felix.hofmann@eng.ox.ac.uk}
\affiliation{
 Department of Engineering Science, University of Oxford, Parks Road, OX1 3PJ Oxford, UK \\
}

\date{\today}

\begin{abstract}
We investigate the mechanisms involved in the thermomigration of interstitial hydrogen in metals. Using irreversible thermodynamics, we develop a comprehensive mechanistic model to capture the controlling effects. Crucially, through validation against published experimental data, our results demonstrate that an electron-wind effect plays a significant role, particularly for materials in which the thermomigration direction matches the heat flux. These findings provide new insights into the factors that affect the localisation of solutes in metals. Moreover, our results indicate that atomistic models may be inadequate for detailed thermomigration studies due to the omission of electronic effects. 
\end{abstract}

\maketitle

In response to the societal shift towards green energy, hydrogen has emerged as a major candidate for energy storage, energy distribution, and transport. Among engineering and scientific communities, there is an associated increasing demand for high-fidelity mechanistic models that can capture hydrogen absorption and redistribution in metals, and ultimately hydrogen embrittlement (HE). Here we consider hydrogen redistribution resulting from temperature gradients (thermomigration) and introduce a new theoretical framework to simulate this effect. Remarkably, no current theory can correctly capture available experimental measurements. However, as concluded by Gonzalez and Oriani \cite{GONZALEZ1965}, any such endeavour would be a `formidable undertaking'. In the literature, discussion of hydrogen redistribution has focused on the effects of hydrostatic stress \cite{LUFRANO1998,MARTINEZPANEDA2016,DEPOVER2019,ELMUKASHFI2020,ABDOLVAND2019,LIU2021,DAS2022,TONDRO2022,TONDRO2023}. Notably, Elmukashfi et al. \cite{ELMUKASHFI2020} presented a constitutive model for coupled hydrogen diffusion problems based on chemical potential gradients. Surprisingly, this has only been used to capture stress-driven hydrogen redistribution, whilst the role of temperature remains underexplored. However, for real technological problems, e.g., in nuclear fusion reactors \cite{MALO2025,DASGUPTA2023,HASHIZUME2011}, and heat exchangers for hydrogen fuel systems \cite{LAPKA2018,PATRAO2024}, thermomigration of hydrogen and other solutes is also a vital consideration. The key knowledge gap lies in understanding the factors that promote the dominance of either effect.

A chemical potential gradient is a generalised force for mass flow \cite{CALLEN1985} in that it encompasses the effects of temperature, stress, and concentration on diffusion. It is defined as the change in Gibbs free energy with respect to the change in atomic hydrogen (H) concentration, $\mu=\frac{\partial G}{\partial C_{\mathrm{L}}}$. An increase in hydrostatic stress yields a reduction in enthalpy of interstitial H (hence a reduction in the Gibbs free energy) as the strain energy associated with lattice distortion due to H is reduced under volumetric expansion. In effect, this promotes a reduction in chemical potential and an associated increase in H solubility \cite{GUEDES2014}. Hence, H diffuses to regions of high hydrostatic stress, such as crack tips. The effect of temperature is not currently as well understood.

In contrast to stress-driven diffusion \cite{BOGKRIS1971,PETERSON1983,KAMMENZIND1998}, there is a major lack of experimental data for the thermomigration of H in metals. This can be attributed in part to the elusive nature of interstitial H, particularly where elevated temperatures are involved \cite{GAULT2021}. The most prominent practical examples of temperature-driven H redistribution are found in Zr alloy cladding materials for nuclear fission applications \cite{VESCHUNOV2016,SUGISAKI1988,MAKI1975}. During in-service conditions, temperature gradients are established radially through the cladding from internal heating and external cooling by circulating water. Water-side oxidation produces H that is absorbed by the material. However, the formation of Zr hydride precipitates adds complexity to this problem and can somewhat obscure the interpretation of results. Specifically, hydrides have been reported to precipitate near the low-temperature end in steady state, while the high-temperature end remains largely hydride-free \cite{NAGASE2005,BECK2012,KAMERMAN2023}. Sawatzky \cite{SAWATZKY1960} provided compelling experimental evidence showing that the concentration of H within the hydride-forming region increases significantly with increasing temperature, before reducing dramatically where the temperature for the terminal solid solubility for precipitation is exceeded, i.e. in the non-hydride-forming region. These results agree with the numerical predictions of Veschunov et al. \cite{VESCHUNOV2016}, who developed phenomenological H transport and precipitation models. However, these lack a physical foundation for thermomigration. In this paper, we present and validate (using experimental data) a thermodynamic model that accounts for numerous independent mechanisms.

From Elmukashfi et al. \cite{ELMUKASHFI2020}, the Gibbs energy per unit volume of a metal, $G$, is given by, $G=\psi-\sigma_{ij}\varepsilon_{ij}+\mu_0\bar{C}_{\mathrm{L}}-TS_{\mathrm{conf}}$, where $\psi$ is the elastic strain energy density and contains the H hydrostatic swelling contribution \cite{CHAPMAN2025}, $\sigma_{ij}$ is stress, $\varepsilon_{ij}$ is strain, $\mu_0$ the standard chemical potential for H, $\bar{C}_{\mathrm{L}}$ the H lattice concentration, $T$ is temperature, and $S_{\mathrm{conf}}$ is the configurational entropy (entropy of mixing). In most H transport models \cite{ELMUKASHFI2020}, the standard chemical potential, $\mu_0$, is assumed constant. According to Kirchheim and Pundt \cite{KIRCHHEIM2014}, $\mu_0$ mainly comprises vibrational and electronic contributions. Recently, de Andres et al. \cite{deANDRES2019} showed using ab-initio modelling that harmonic vibrations play a crucial role in H transport. For non-isothermal conditions therefore, it is obvious that the constant $\mu_0$ assumption does not hold. Treating $\mu_0\bar{C}_{\mathrm{L}}$ as a function of vibrational and electronic contributions only, we may write

\begin{equation}\label{G2}
    G=\psi-\sigma_{ij}\varepsilon_{ij}+G_{\mathrm{vib}}+G_{\mathrm{e}}-TS_{\mathrm{conf}},\\
\end{equation}

\noindent where $G_{\mathrm{vib}}$ is the vibrational contribution to the Gibbs free energy and $G_{\mathrm{e}}$ is the electronic contribution. Expressing the vibrational contribution as $G_{\mathrm{vib}}=H_{\mathrm{vib}}-TS_{\mathrm{vib}}$, where $H_{\mathrm{vib}}$ and $S_{\mathrm{vib}}$ are the vibrational enthalpy and entropy, respectively, which are obtained from the vibrational partition function \cite{FEDEROV2021}. As the driving force for H transport is related to the chemical potential, we wish to extract the vibrational contribution, $\mu_{\mathrm{vib}}=\frac{\partial G_{\mathrm{vib}}}{\partial\bar{C}_{\mathrm{L}}}=h_{\mathrm{vib}}-Ts_{\mathrm{vib}}$, where $h_{\mathrm{vib}}$ and $s_{\mathrm{vib}}$ are the specific vibrational enthalpy and entropy for interstitial H, respectively.

The electronic contribution to the Gibbs free energy is given in terms of an electrical potential, $G_{\mathrm{e}}=-F\phi\bar{C}_{\mathrm{e}}$ \cite{CAMPERO2020}, where $F$ is the Faraday constant, $\phi$ the electrical potential, and $\bar{C}_{\mathrm{e}}$ is the molar concentration of free electrons. Based on the electrostatic transfer of charge from the lattice to interstitial H in an open circuit, we may write the electronic chemical potential for H transport as $\mu_{\mathrm{e}}=\frac{\partial G_{\mathrm{e}}}{\partial\bar{C}_{\mathrm{L}}}=\frac{\partial\bar{C}_{\mathrm{e}}}{\partial\bar{C}_{\mathrm{L}}}\frac{\partial G_{\mathrm{e}}}{\partial\bar{C}_{\mathrm{e}}}=-Z_{\mathrm{es}}F\phi$, where $Z_{\mathrm{es}}=\frac{\partial\bar{C}_{\mathrm{e}}}{\partial\bar{C}_{\mathrm{L}}}$ is the charge transfer coefficient. For non-isothermal problems in metals, we consider thermoelectric fields that arise according to $\phi=\alpha_{\mathrm{S}}T$ and $\nabla\phi=T\frac{\partial\alpha_{\mathrm{S}}}{\partial T}\nabla T+\alpha_{\mathrm{S}}\nabla T=(\alpha_{\mathrm{T}}+\alpha_{\mathrm{S}})\nabla T$, where $\alpha_{\mathrm{S}}$ is the Seebeck coefficient (which relates the induced thermoelectric voltage to the temperature difference across a material) and $\alpha_{\mathrm{T}}=T\frac{\partial\alpha_{\mathrm{S}}}{\partial T}$ is the Thomson coefficient (note that the Thomson effect may be disregarded for materials where the thermoelectric power is independent of temperature).

As shown in \cite{ELMUKASHFI2020}, the hydrostatic stress contribution to the chemical potential derives from hydrogen-induced volumetric lattice swelling, since the elastic strain energy density, $\psi=\frac{1}{2}(\varepsilon_{ij}-\varepsilon_{ij}^{\mathrm{pl}}-\varepsilon_{ij}^{\mathrm{s}})\mathbb{C}_{ijkl}(\varepsilon_{kl}-\varepsilon_{kl}^{\mathrm{pl}}-\varepsilon_{kl}^{\mathrm{s}})$ and the swelling strain is given by $\varepsilon_{ij}^{\mathrm{s}}=\frac{1}{3}V_{\mathrm{L}}\bar{C}_{\mathrm{L}}\delta_{ij}$, where $V_{\mathrm{L}}$ is the partial molar volume of H in the solvent lattice and $\bar{C}_{\mathrm{L}}$ the interstitial hydrogen concentration. Hence, $\frac{\partial\psi}{\partial\bar{C}_{\mathrm{L}}}=-V_{\mathrm{L}}\sigma_{\mathrm{H}}$, where $\sigma_{\mathrm{H}}=\frac{1}{3}\sigma_{kk}$ is the hydrostatic stress. Furthermore, the configurational entropy contribution to Equation (\ref{G2}) yields the widely reported logarithmic relationship between the chemical potential and H concentration, $\mu_{\mathrm{conf}}=\frac{\partial}{\partial \bar{C}_{\mathrm{L}}}(-TS_{\mathrm{conf}})=RT\ln{\left(\frac{\bar{C}_{\mathrm{L}}}{\bar{C}_{\mathrm{L}}^{\mathrm{max}}-\bar{C}_{\mathrm{L}}}\right)}$, since $\frac{\partial S_{\mathrm{conf}}}{\partial\bar{C}_{\mathrm{L}}}=-R\ln{\left(\frac{\bar{C}_{\mathrm{L}}}{\bar{C}_{\mathrm{L}}^{\mathrm{max}}-\bar{C}_{\mathrm{L}}}\right)}$. For low lattice concentrations relevant to metals, we may write the lattice occupancy, $\theta_{\mathrm{L}}=\frac{\bar{C}_{\mathrm{L}}}{\bar{C}_{\mathrm{L}}^{\mathrm{max}}}\approx\frac{\bar{C}_{\mathrm{L}}}{\bar{C}_{\mathrm{L}}^{\mathrm{max}}-\bar{C}_{\mathrm{L}}}$, where $\bar{C}_{\mathrm{L}}^{\mathrm{max}}$ represents the theoretical maximum interstitial H concentration (corresponding to the total number of interstitial sites). The modified chemical potential is now given by

\begin{equation}\label{mu}
    \mu=-V_{\mathrm{L}}\sigma_{\mathrm{H}}+h_{\mathrm{vib}}-Ts_{\mathrm{vib}}-Z_{\mathrm{es}} F\alpha_{\mathrm{S}}T+RT\ln{(\theta_{\mathrm{L}})}.\\
\end{equation}

\noindent In H transport modelling \cite{ZHANG2022}, the heat of transport, $Q^*$, has been used as a phenomenological term to capture thermomigration. Zhang et al. \cite{ZHANG2022} presented a modified chemical potential-based model for H transport in Ni alloys under temperature and hydrostatic stress gradients; the H flux was given by $\bar{\boldsymbol{J}}_{\mathrm{H}}=-\frac{D_{\mathrm{L}}\bar{C}_{\mathrm{L}}}{RT}\left(-V_{\mathrm{L}}\nabla\sigma_{\mathrm{H}}+\frac{Q^*}{T}\nabla T+RT\nabla\ln{\theta_{\mathrm{L}}}\right)$, where $D_{\mathrm{L}}$ is H diffusivity. Gradients of temperature, stress, and lattice H concentration were treated as explicit driving forces for diffusion. In Longhurst's model \cite{LONGHURST1985}, the chemical potential and temperature gradients were treated as distinct driving forces, which is peculiar, given that the chemical potential is clearly a function of temperature. Here, we use our holistic description of the H chemical potential to derive an expression for $Q^*$. From irreversible thermodynamics \cite{ONSAGER1931I,ONSAGER1931II}, we write coupled equations for hydrogen flux, $\boldsymbol{\bar{J}}_{\mathrm{H}}$, and heat flux, ${\boldsymbol{J}}_{\mathrm{q}}$, as 

\begin{equation}\label{Onsager}
    \begin{pmatrix}
    \boldsymbol{\bar{J}}_{\mathrm{H}}\\
    {\boldsymbol{J}}_{\mathrm{q}}
    \end{pmatrix}=
    \begin{pmatrix}
    L_{\mathrm{HH}} & L_{\mathrm{Hq}}\\
    L_{\mathrm{qH}} & L_{\mathrm{qq}}
    \end{pmatrix}
    \begin{pmatrix}
    \boldsymbol{X}_{\mathrm{H}}\\
    \boldsymbol{X}_{\mathrm{q}}
    \end{pmatrix}\\,
\end{equation}

\noindent where the thermodynamic driving forces are defined as $\boldsymbol{X}_{\mathrm{H}} = -\nabla\left(\frac{\mu}{T}\right)$ and $\boldsymbol{X}_{\mathrm{q}} = \nabla\left(\frac{1}{T}\right)$, and the coefficients $L_{ij}$ are the Onsager transport coefficients. In irreversible thermodynamics, the driving force for mass transport, $\boldsymbol{X}_{\mathrm{H}}$, arises from the principle of entropy production, rather than from energy minimisation \cite{DEGROOT2013}. Consequently, the commonly used isothermal diffusion law based solely on chemical potential gradients,
$\boldsymbol{\bar{J}}_{\mathrm{H}} = -\frac{D_{\mathrm{L}}\bar{C}_{\mathrm{L}}}{RT}\nabla\mu$,
is insufficient for describing non-isothermal systems. Onsager's reciprocity condition tells us that the off-diagonal terms, $L_{\mathrm{Hq}}$ and $L_{\mathrm{qH}}$, are equal. In the first instance, consider the situation where $L_{\mathrm{Hq}}=L_{\mathrm{qH}}=0$: we then have from Equation (\ref{Onsager}),
\begin{subequations}\label{Onsager_zero_offdiag}
\begin{align}
\quad
\boldsymbol{\bar{J}}_{\mathrm{H}}
&= L_{\mathrm{HH}}\Big(-\frac{1}{T}\nabla\mu+\frac{\mu}{T^{2}}\nabla T\Big),\\
\quad
\boldsymbol{J}_{\mathrm{q}}
&= -\frac{L_{\mathrm{qq}}}{T^{2}}\,\nabla T,
\end{align}
\end{subequations}

\noindent where $L_{\mathrm{HH}}=\frac{D_{\mathrm{L}}\bar{C}_{\mathrm{L}}}{R}$ and $L_{\mathrm{qq}}=\kappa T^2$ and $\kappa$ is thermal conductivity. Note that for isothermal problems, Equation (\ref{Onsager_zero_offdiag}a) reduces to $\bar{\boldsymbol{J}}_{\mathrm{H}}=-\frac{D_{\mathrm{L}}\bar{C}_{\mathrm{L}}}{RT}\nabla\mu$. From Equation (\ref{mu}), we may write the chemical potential gradient as a function of the spatially varying fields as 

\begin{equation}\label{gradmu0}
    \nabla\mu=-V_{\mathrm{L}}\nabla\sigma_{\mathrm{H}}-s_{\mathrm{vib}}\nabla T-Z_{\mathrm{es}}F\left(\alpha_{\mathrm{S}}+\alpha_{\mathrm{T}}\right)\nabla T+R\ln{\left(\theta_{\mathrm{L}}\right)}\nabla T+RT\nabla\ln{\left(\theta_{\mathrm{L}}\right)}.\\
\end{equation}

\noindent The vibrational enthalpy and entropy terms are derived from the vibrational partition function \cite{FEDEROV2021} and their temperature derivatives can be shown to yield $T\frac{\partial h_{\mathrm{vib}}}{\partial T}=T^2\frac{\partial s_{\mathrm{vib}}}{\partial T}$, which eliminates the dependence of $\nabla\mu$ on $h_{\mathrm{vib}}$. Combining Equations (\ref{mu}), (\ref{Onsager_zero_offdiag}a), and (\ref{gradmu0}) yields 

\begin{equation}\label{JH}
    \boldsymbol{\bar{J}}_{\mathrm{H}}=-\frac{L_{\mathrm{HH}}}{T}\left(-V_{\mathrm{L}}\nabla\sigma_{\mathrm{H}}+\left(V_{\mathrm{L}}\sigma_{\mathrm{H}}-h_{\mathrm{vib}}-Z_{\mathrm{es}}F\alpha_{\mathrm{T}}T\right)\frac{\nabla T}{T}+RT\nabla\ln{\left(\theta_{\mathrm{L}}\right)}\right).\\
\end{equation}

\noindent To derive $Q^*$ from our current model for H transport, we may equate Equation (\ref{JH}) to a general model for hydrogen transport, e.g., from Zhang et al. \cite{ZHANG2022}, where $\boldsymbol{\bar{J}}_{\mathrm{H}}=-\frac{L_{\mathrm{HH}}}{T}\left(-V_{\mathrm{L}}\nabla\sigma_{\mathrm{H}}+\frac{Q^*}{T}\nabla T+RT\nabla\ln{\theta_{\mathrm{L}}}\right)$. $Q^*$ is hence given by

\begin{equation}\label{Q*1}
    Q^*=V_{\mathrm{L}}\sigma_{\mathrm{H}}-h_{\mathrm{vib}}-Z_{\mathrm{es}}F\alpha_{\mathrm{T}}T.
\end{equation}

\noindent The vibrational enthalpy per mole is

\begin{equation}\label{h}
    h_{\mathrm{vib}}=3N_{\mathrm{A}}\hbar\omega\left(1+\left(\exp{\left(\frac{\hbar\omega}{k_{\mathrm{B}}T}\right)}-1\right)^{-1}\right),\\
\end{equation}

\noindent where $N_{\mathrm{A}}$ is Avogadro's constant, $\hbar$ is Planck's constant, $\omega$ the H atom vibrational frequency, and $k_{\mathrm{B}}$ is Boltzmann's constant \cite{DASILVA1976}. The vibrational frequency depends on the interatomic potentials and masses of H and of the solvent atom \cite{BATES1928}. Because of the vast differences between the vibrational frequencies of H and the host metal, it is assumed that the H vibrational entropy is independent of the latter. 

Here we show that where $L_{\mathrm{Hq}}=L_{\mathrm{qH}}=0$, the heat of transport is given by the sum of distinct contributions arising from: hydrostatic stress, vibrational enthalpy, and electrostatic interactions. According to Wipf \cite{WIPF2005}, the primary contributions to $Q^*$ originate from (i) an intrinsic hopping rate difference between high and low temperature regions (resulting in a net flux across atomic planes), (ii) thermoelectric fields, (iii) electron scattering, and (iv) phonon scattering. Equation (\ref{Q*1}) consists of (i) intrinsic, $Q^*_{\mathrm{int}}=V_{\mathrm{L}}\sigma_{\mathrm{H}}-h_{\mathrm{vib}}$, and (ii) electrostatic, $Q^*_{\mathrm{es}}=-Z_{\mathrm{es}}FT\alpha_{\mathrm{T}}$, factors, both of which were derived from our improved description of the Gibbs energy (Equation (\ref{G2})). Our intrinsic contribution bears a resemblance to the classical Wirtz model for vacancy thermomigration (from transition state theory) \cite{WIRTZ1943}. In Wirtz’s formulation, the driving force for vacancy transport arises from the difference between the migration and formation enthalpies of vacancies. Both quantities depend sensitively on the local hydrostatic stress and the vibrational enthalpy. However, stress was not previously reported to affect thermomigration. Using Equation (\ref{Q*1}), a rough estimate of its effect in Fe, where $V_{\mathrm{L}}\approx4.1$ {\AA}$^3$/atom \cite{ZHANG2024}, suggests that it yields a comparatively small contribution to $Q^*$ on the order of 2.5 J/mol per 1 MPa. Our electrostatic model, (ii), is consistent with that of Fiks \cite{FIKS1961,HUNTINGTON1968}. Factors (iii) and (iv) relate to thermal transport in metals, and hence, associated H fluxes should only derive from $L_{\mathrm{Hq}}\boldsymbol{X}_{\mathrm{q}}$. These mechanisms will be discussed in greater detail later. As there is no available experimental evidence of H-affected thermal transport, we will first consider Equation (\ref{Q*1}) alone, such that $L_{\mathrm{Hq}}=L_{\mathrm{qH}}=0$.

\begin{figure}[ht]
\includegraphics[width=0.5\linewidth]{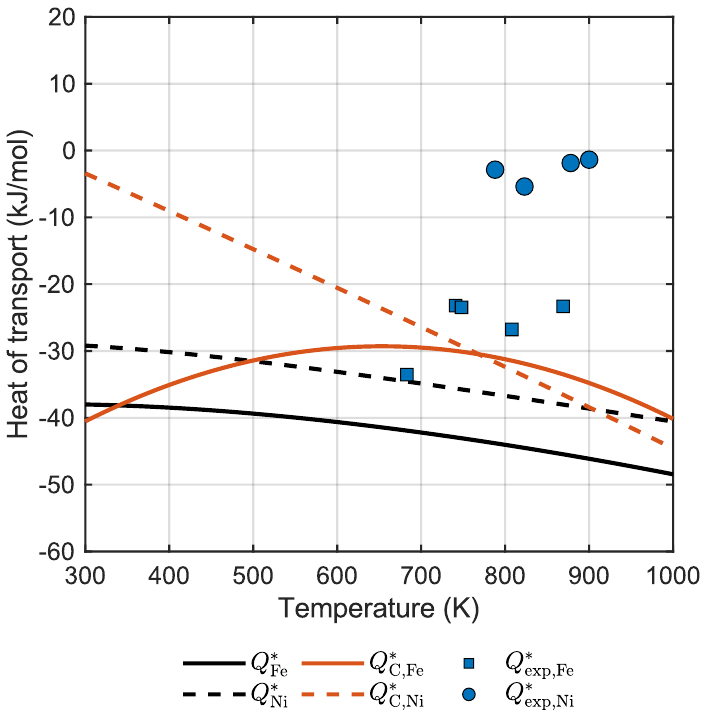}
\caption{\label{CALPHAD_comparison} A comparison of the predicted temperature-dependence of the heat of transport using our mechanistic model derived from the $\boldsymbol{X}_{\mathrm{H}}$ driving force (shown in black) and the CALPHAD-based model (shown in red) with experimental data from Gonzalez and Oriani \cite{GONZALEZ1965} for Fe and Ni. Continuous lines represent predictions in Fe and dashed lines represent predictions in Ni. }
\end{figure}

As discussed earlier, very little heat of transport data is available for non-hydride-forming materials. To our knowledge, the only published experimental measurements of $Q^*$ for $\alpha$-Fe and Ni (materials of interest to the HE community) were by Gonzalez and Oriani \cite{GONZALEZ1965} in the 1960's. The authors used a thermo-osmosis technique to measure $Q^*$ in the temperature range $400 - 600$ $^\mathrm{o}$C, which involved enforcing a temperature gradient across each material (with temperature control at hot and cold ends, $T_{\mathrm{h}}$ and $T_{\mathrm{c}}$, respectively), while applying a hydrogen gas pressure, $p_{\mathrm{h}}$, at the hot end, and measuring the resultant cold end gas pressure, $p_{\mathrm{c}}$, at steady state. Elevated temperatures were necessary for this procedure to ensure that transients were diffusion-controlled rather than surface-controlled (as the steady state pressure is determined by interpolation between positive and negative fluxes). Using the relationship, $Q^*_{\mathrm{exp}}+\Delta H=-\frac{RT_{\mathrm{ave}}^2\ln{(\frac{p_{\mathrm{c}}}{p_{\mathrm{h}}})}}{2(T_{\mathrm{c}}-T_{\mathrm{h}})}$, heat of transport measurements were obtained, where $\Delta H$ is the heat of solution and $T_{\mathrm{ave}}$ the average temperature. The heat of solution must therefore be known to extract $Q^*$ using this procedure. The authors assumed a fixed value for $\Delta H$ for each material. Whilst all measurements in Fe and Ni were negative over the temperature range studied (implying redistribution from cold to hot along temperature gradients), $Q^*$ was shown on average to increase with increasing temperature. For most hydride-forming materials (Zr, Ti, Nb), which are central to most H thermomigration studies, $Q^*$ is reported as a positive value \cite{ORIANI1969}, i.e., H flux is of the same sign as heat flux. To avoid the complexity associated with hydride formation, however, the experimental data sets for hydride-free Fe and Ni \cite{GONZALEZ1965} are used to evaluate our new model. In experiments, the vibrational frequency, $\omega$, is often given in terms of the wave number \cite{KOLESOV2017}, $\bar{\omega}=\omega/c$, where $c$ is the speed of light in a vacuum. The wave numbers used here for Fe and Ni are given in Table \ref{properties}. For hydrogen at interstitial lattice sites, charge transfer has been estimated using density functional theory (DFT) in a number of studies. Juan and Hoffmann~\cite{JUAN1999} reported charge transfer, $Z_{\mathrm{es}}$, of 0.12~$e$ and 0.16~$e$ (where $e$ is the elementary charge, $1.602\times10^{-19}\,\mathrm{C}$) from Fe to H in tetrahedral and octahedral lattice sites, respectively, while Itsumi and Ellis~\cite{ITSUMI1996} reported charge transfer of up to 0.6~$e$. In Ni, Weng et al. \cite{WENG2012} reported a charge transfer of 0.25 $e$. Charge transfer properties used in this study are also given in Table \ref{properties}. In Figure \ref{CALPHAD_comparison}, we present a comparison of the predictions obtained from Equation (\ref{Q*1}) with those from CALPHAD Gibbs free energy functions \cite{PERRUT2015}. 

The CALPHAD Gibbs free energy functions for the Fe-H and Ni-H systems originate from experimental thermodynamic data and first-principles calculations with a functional form dependent on temperature expressed as $G_{\mathrm{C}}=a+bT+cT\ln{T}+dT^2+eT^{-1}+fT^3$. The coefficients $a$ to $f$ are fitting parameters and are dependent on H concentration. These functions are utilised here to derive temperature-dependent heat of transport distributions from $Q^*_{\mathrm{C}}=T\frac{\partial}{\partial T}\left(\frac{\partial G_{\mathrm{C}}}{\partial\bar{C}_{\mathrm{L}}}\right)-\frac{\partial G_{\mathrm{C}}}{\partial\bar{C}_{\mathrm{L}}}$, since Equations (\ref{Onsager_zero_offdiag}a) and (\ref{JH}) can be shown to give $Q^*=T\frac{\partial\mu}{\partial T}-\mu$. Figure \ref{CALPHAD_comparison} shows weak agreement between Equation (\ref{Q*1}) and CALPHAD-predicted heat of transport distributions, but qualitative similarities, in that the gradient of $Q^*$ with respect to $T$ is mostly negative. In contrast, both models reveal striking disagreement with Gonzalez and Oriani's \cite{GONZALEZ1965} experimental data. Having carefully considered key contributors to the H chemical potential, we therefore hypothesise that these major discrepancies arise from a non-zero off-diagonal Onsager coefficient, $L_{\mathrm{Hq}}$. Specifically, factors (iii) electron scattering, and/or (iv) phonon scattering, are implicated, since Onsager reciprocity implies that the absent $Q^*$ contributions have a direct effect on heat carriers (electrons and phonons). In each case, the driving force derives from momentum transfer via H impurity scattering \cite{WIPF2005}, which biases the H jump direction at the saddle point. The role of phonon scattering in momentum transfer is a subject of controversy, particularly as it requires that phonon motion be considered independent of particle motion, and that phonons transfer real momentum \cite{HUNTINGTON1968}. For these reasons, we do not consider (iv). The role of electron scattering, however, is said to give rise to a considerable contribution to $Q^*$ in metals with both electron and hole bands \cite{WIPF2005}. In an open-circuit, a zero net electric current is imposed, such that the only contribution from (iii) in metals with a single electron band originates from differences in momentum and scattering between electrons at different temperatures (and is likely to be small) \cite{WIPF2005}. In metals with electron and hole bands (such as Fe and Ni), however, isodirectional fluxes of both electrons and holes can arise in the direction of heat flux while maintaining a zero net electric current \cite{WIPF2005}. This forms the basis for Huntington's model for the charge carrier scattering contribution to the heat of transport \cite{HUNTINGTON1968}, $Q^*_{\mathrm{ele}}$, sometimes called an electron-wind, which we employ here. The model was derived by considering the electron drag force for a single ion, $\boldsymbol{F}_{\mathrm{ele}}
= \frac{1}{4\pi^3} \int_{\mathbb{R}^3} \hbar\boldsymbol{k}\, v_{\mathrm{F}}\, \varsigma_{\mathrm{H}}\, f(\boldsymbol{k})\, \mathrm{d}^3\boldsymbol{k}
= -\frac{Q^*_{\mathrm{ele}}}{N_{\mathrm{A}}T}\nabla T$
where $\boldsymbol{k}$ is the electron wavevector (so that $\hbar\boldsymbol{k}$ is the electron crystal momentum), 
$f(\boldsymbol{k})$ is the electron distribution function, $v_{\mathrm{F}}$ the electron velocity at the Fermi surface, and $\varsigma_{\mathrm{H}}$ the ion (H) scattering cross-section for electrons. The integral is taken over all $\boldsymbol{k}$ in reciprocal (momentum) space. The distribution function, $f$, includes equilibrium and non-equilibrium parts, whose volume integrals are characterised by terms $K_{\mathrm{F}0}$ and $K_{\mathrm{F}1}$, respectively. An analogous expression for the electric current was also derived, with equilibrium and non-equilibrium integrals, $K_{\mathrm{J}0}$ and $K_{\mathrm{J}1}$. The resulting $Q^*_{\mathrm{ele}}$ is given by 

\begin{equation}\label{Q*ele1}
    Q^*_{\mathrm{ele}}=-FT\alpha_{\mathrm{S}}\frac{1+\gamma}{1-2\gamma}\left(1+3\delta\right)\varsigma_{\mathrm{H}}\lambda_{\mathrm{e}}N_{\mathrm{e}},\\
\end{equation}

\noindent where $\gamma$ and $\delta$ are dimensionless quantities which represent the ratio of the hole to electron band contribution to $K_{\mathrm{J}0}$ and $K_{\mathrm{F}0}$, respectively. $\lambda_{\mathrm{e}}$ is the electron mean free path, and $N_{\mathrm{e}}$ is the free electron density. The difficulty here lies in quantifying $\gamma$ and $\delta$. On the basis that the number density of both charge carriers are equal, $\gamma$ is approximated as the ratio of the electron effective mass to hole effective mass (since $K_{\mathrm{J}0}\propto\frac{1}{m^*}$), giving $\gamma\approx0.1$ \cite{HUNTINGTON1968}. For $\delta$, experimental measurements for the charge of transport, $Z^*$, are used to infer approximate values for Fe and Ni; for most one band metals, $Z^*$ is on the order of 1.6 $e$, while it ranges between just 0.24 to 0.28 $e$ in Fe \cite{ORIANI1976,ORIANI1967EM} and 0.57 to 0.67 $e$ in Ni \cite{ORIANI1976,ORIANI1967EM}. In Fe, for example, we approximate $\delta\approx\frac{1.6-0.26}{1.6}=0.84$, which closely aligns with \cite{HUNTINGTON1968}. In Ni, we estimate $\delta\approx0.61$ using the same approach. In Equation (\ref{Q*ele1}), it is interesting to note the dependence of $Q^*_{\mathrm{ele}}$ on $\alpha_{\mathrm{S}}$; this is derived from a dependence on the electron energy derivative of scattering time (rather than from electrostatic fields, as in $\nabla\mu$). To enhance Huntington's model \cite{HUNTINGTON1968}, we relate the product $\varsigma_{\mathrm{e}}\lambda_{\mathrm{e}}$ to further experimental observables. From Matthiessen's rule, we write the total electrical resistivity as the sum of temperature-dependent resistivity and resistivity due to H impurity scattering, $\rho=\rho_{\mathrm{th}}+\rho_{\mathrm{H}}$, where $\rho_{i}=\frac{m_{\mathrm{e}}}{N_{\mathrm{e}}e^2\tau_{i}}$, $m_{\mathrm{e}}$ is electron mass, and $\tau_{i}$ represents thermal or H impurity scattering time. From Fiks \cite{FIKS1961}, $\tau_{\mathrm{H}}^{-1}=N_{\mathrm{H}}v_{\mathrm{F}}\varsigma_{\mathrm{H}}$, where $N_{\mathrm{H}}$ is the H number density. When combined with the expression for $\rho_{\mathrm{H}}$, this yields $\varsigma_{\mathrm{H}}=\frac{d\rho_{\mathrm{H}}}{dN_{\mathrm{H}}}\cdot\frac{N_{\mathrm{e}}e^2}{m_{\mathrm{e}}v_{\mathrm{F}}}=\frac{d\rho}{dN_{\mathrm{H}}}\cdot\frac{N_{\mathrm{e}}e^2}{m_{\mathrm{e}}v_{\mathrm{F}}}$. Assuming the electron mean free path is dominated by thermal scattering, we approximate $\lambda_{\mathrm{e}}\approx\tau_{\mathrm{th}}v_{\mathrm{F}}$, which yields $\varsigma_{\mathrm{H}}\lambda_{\mathrm{e}}N_{\mathrm{e}}=\frac{d\rho}{dN_{\mathrm{H}}}\cdot\frac{N_{\mathrm{e}}}{\rho_{\mathrm{th}}}$. Hence, our revised model for the electron-wind effect is given by

\begin{equation}\label{Q*ele2}
    Q^*_{\mathrm{ele}}=-FT\alpha_{\mathrm{S}}\frac{1+\gamma}{1-2\gamma}\left(1+3\delta\right)\frac{d\rho}{dN_{\mathrm{H}}}\cdot\frac{N_{\mathrm{e}}}{\rho_{\mathrm{th}}}.\\
\end{equation}

\noindent In the low concentration limit (as is appropriate for H in metals), $\frac{d\rho}{dN_{\mathrm{H}}}$ is treated as constant, while polynomial fitting is used to characterise the temperature dependence of $\rho_{\mathrm{th}}$ in the temperature range 300 K to 1000 K. Polynomial fitting is also used to capture temperature-dependent Seebeck coefficient data \cite{SECCO2017,HAUPT2020}, and associated Thomson coefficient data. For both Fe and Ni, $N_{\mathrm{e}}=2N_{\mathrm{L}}$, where $N_{\mathrm{L}}$ is the number of solvent atoms per unit volume, treated as a constant. A summary of property data is given in Table \ref{properties}. Accounting for the electron-wind effect, the heat of transport (Equation (\ref{Q*1})) is now

\begin{equation}\label{Q*2}
    Q^*=V_{\mathrm{L}}\sigma_{\mathrm{H}}-h_{\mathrm{vib}}-Z_{\mathrm{es}}F\alpha_{\mathrm{T}}T-FT\alpha_{\mathrm{S}}\frac{1+\gamma}{1-2\gamma}\left(1+3\delta\right)\frac{d\rho}{dN_{\mathrm{H}}}\cdot\frac{N_{\mathrm{e}}}{\rho_{\mathrm{th}}}.\\
\end{equation}

\begin{figure}[ht]
\includegraphics[width=1\linewidth]{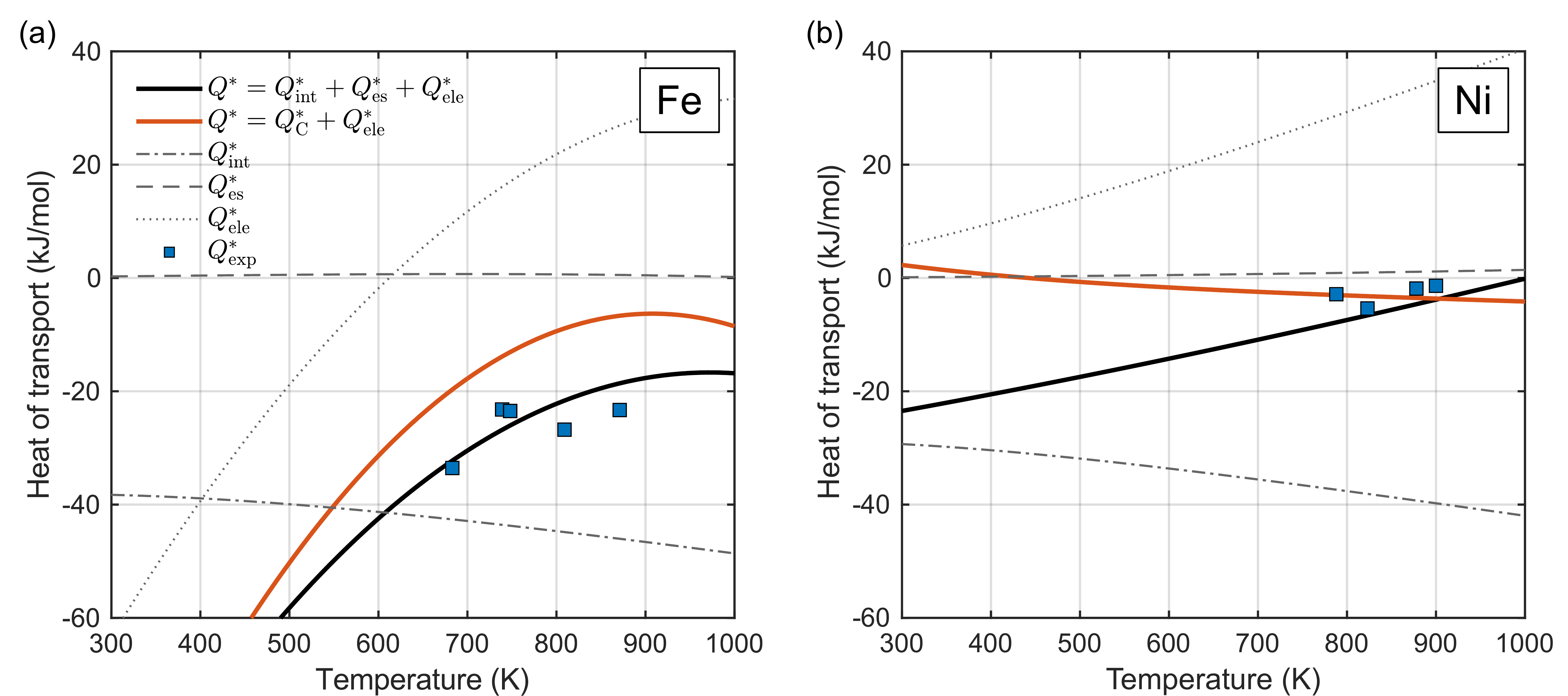}
\caption{\label{Qstar_comparison} Validation of the revised heat of transport model, accounting for the electron-wind effect by comparison with experimental data from Gonzalez and Oriani \cite{GONZALEZ1965} for (a) Fe and (b) Ni. Dashed lines represent individual contributions to the total heat of transport.}
\end{figure}

\noindent Maintaining consistency with the earlier phenomenological model for hydrogen flux \cite{ZHANG2019}, $\bar{\boldsymbol{J}}_{\mathrm{H}}=-\frac{D_{\mathrm{L}}\bar{C}_{\mathrm{L}}}{RT}\left(-V_{\mathrm{L}}\nabla\sigma_{\mathrm{H}}+\frac{Q^*}{T}\nabla T+RT\nabla\ln{\theta_{\mathrm{L}}}\right)$, where $Q^*=Q^*_{\mathrm{int}}+Q^*_{\mathrm{es}}+Q^*_{\mathrm{ele}}$, we may write $L_{\mathrm{Hq}}\boldsymbol{X}_{\mathrm{q}}=-\frac{D_{\mathrm{L}}\bar{C}_{\mathrm{L}}}{RT}\cdot\frac{Q^*_{\mathrm{ele}}}{T}\nabla T$. Hence, the off-diagonal Onsager coefficients are now given by $L_{\mathrm{Hq}}=L_{\mathrm{qH}}=L_{\mathrm{HH}}Q^*_{\mathrm{ele}}$. A comparison of this revised model with experimental data is presented in Figure \ref{Qstar_comparison}, showing a substantial improvement over Equation (\ref{Q*1}). 

Using experimental thermoelectric and resistivity data (see Table \ref{properties}), the electron-wind contribution is shown to significantly alter the functional form of $Q^*$, so that the experimental trends \cite{GONZALEZ1965} are captured; crucially, no free variables or fitting parameters were required in doing so. These results support our earlier hypothesis, that thermal transport mechanisms play a key role in H thermomigration. The individual contributions to Equation (\ref{Q*2}) and a modified CALPHAD-based model, $Q^*=Q^*_{\mathrm{C}}+Q^*_{\mathrm{ele}}$, are also shown. Notably, the electrostatic contribution to $\boldsymbol{X}_{\mathrm{H}}$ is shown to be negligible for each material. For solutes with larger effective charge, $Z_{\mathrm{es}}$, however, $Q^*_{\mathrm{es}}$ may represent a more important contribution. As the intrinsic contribution is given by $-h_{\mathrm{vib}}$ (where $\sigma_{\mathrm{H}}=0$), it will always promote $Q^*<0$ and $\frac{\partial Q^*}{\partial T}<0$. The electron-wind contribution, however, is shown to give rise to $\frac{\partial Q^*}{\partial T}>0$ in Fe and Ni, consistent with experimental observations. This suggests that in positive $Q^*$ materials like Zr \cite{KANG2023}, electron-wind effects are dominant. In the absence of reliable thermoelectric data, however, this is yet to be confirmed. In Ni, Equation (\ref{Q*2}) appears to slightly under-predict the experimental data. It is worth noting that this discrepancy may arise due to the omission of anharmonic contributions to $h_{\mathrm{vib}}$, which may only be investigated using high-fidelity atomistic modelling. Interestingly, the CALPHAD-based model for $Q^*$ with electron-wind correction is shown to yield improved agreement with the experimental data, though its temperature dependence is inconsistent with our mechanistic model. Nevertheless, we demonstrate its potential to aid in deriving rough estimates for $Q^*$. 

\begin{table}[h]%The best place to locate the table environment is directly after its first reference in text
\caption{\label{properties}%
Property data used in modelling the heat of transport. Where data are obtained from experimental measurements, references are provided. 
}
\begin{ruledtabular}
\begin{tabular}{ccccc}
\textrm{Property}&
\textrm{Fe}&
\textrm{Ni}&
\textrm{Units}&
\textrm{Source}\\
\colrule
$\bar{\omega}$ & 1060 & 800 & cm$^{-1}$ & \cite{baro1981,HOCHARD1995} \\
$Z_{\mathrm{es}}$ & 0.6 & 0.25 & $e$ & \cite{ITSUMI1996,WENG2012} \\
$\alpha_{\mathrm{S}}$ & 3.58$\times$10$^{-5}T^2$-0.07$T$+32.2 & -0.03$T$+1.74 & $\mu$VK$^{-1}$ & \cite{SECCO2017,HAUPT2020} \\
$\alpha_{\mathrm{T}}$ & 7.14$\times$10$^{-5}T^2$-0.07$T$ & -0.03$T$ & $\mu$VK$^{-1}$ & $T\frac{\partial\alpha_{\mathrm{S}}}{\partial T}$ \\
$\rho_{\mathrm{th}}$ & 4.00$\times$10$^{-10}T$ & 2.26$\times$10$^{-10}T+$1.20$\times$10$^{-7}$ & $\Omega$m & \cite{WAGENKNECHT2015} \\
$\frac{d\rho}{dN_{\mathrm{H}}}$ & 2.45$\times$10$^{-35}$ & 7.31$\times$10$^{-36}$ & $\Omega$m$^4$ & \cite{SINGH2025,PAPASTAIKOUDIS1983} \\
$N_{\mathrm{L}}$ & 8.49$\times$10$^{28}$ & 9.15$\times$10$^{28}$ & m$^{-3}$ & \cite{SMITHELLS} \\
\end{tabular}
\end{ruledtabular}
\end{table}

In light of these findings, an interesting discussion emerges from the failure of CALPHAD models alone to capture $Q^*$ measurements. In particular, parallels between associated thermodynamic data, that is, data excluding the electron-wind effect, and molecular dynamics (MD) modelling can be made. Given that classical MD neglects electronic contributions entirely, the methodology is also likely to be an unreliable predictor of thermomigration in metals. This highlights a crucial need to consider the electron-wind effect when modelling such problems.

\section*{Acknowledgements}

\noindent The authors would like to acknowledge Rolls-Royce plc. for their financial and technical support in this project (grant number RR/UTC/89/9 BPC 189). We would particularly like to thank Chris Argyrakis, Louise Gale, and Duncan Maclachlan for their input. We would also like to acknowledge Daniel Mason, from the UK Atomic Energy Authority, for his insight and thought-provoking discussions on this topic. 

\bibliographystyle{unsrt}
\bibliography{References}% Produces the bibliography via BibTeX.

\end{document}